\documentclass[aps,pra,reprint,superscriptaddress]{revtex4-1}

\usepackage{amsmath}
\usepackage{amsfonts}
\usepackage{amssymb}
\usepackage{graphicx}
\usepackage{graphics}

\begin{document}


\title{A surface electrode point Paul trap}


\author{Tony Hyun Kim}
\email{kimt@mit.edu}
\author{Peter F. Herskind}
\affiliation{Center for Ultracold Atoms, Department of Physics, Massachusetts Institute of Technology\\
77 Massachusetts Avenue, Cambridge, MA 02139}
\author{Taehyun Kim}
\author{Jungsang Kim}
\affiliation{Department of Electrical and Computer Engineering, Duke University Durham, NC 27708}
\author{Isaac L. Chuang$^1$}

\date{\today}

\begin{abstract}
We present a model as well as experimental results for a surface electrode radio-frequency Paul trap that has a circular electrode geometry well-suited for trapping of single ions and two-dimensional planar ion crystals. The trap design is compatible with microfabrication and offers a simple method by which the height of the trapped ions above the surface may be changed \emph{in situ}. We demonstrate trapping of single $^{88}$Sr$^+$ ions over an ion height range of $200$--$1000\mu$m for several hours under Doppler laser cooling, and use these to characterize the trap, finding good agreement with our model.
\end{abstract}

\pacs{}

\maketitle

\section{Introduction \label{Introduction}}
Radiofrequency (rf) traps have been applied extensively in a large variety of scientific studies over the past six decades. Originating from mass spectrometry~\cite{Paul1953}, they have then been applied in fields such as metrology~\cite{Rosenband2008}, quantum information science~\cite{Nielsen2000,Blatt2008} and cold molecular physics~\cite{Staanum2010,Schneider2010}, to mention but a few.

\begin{figure}[b]
\includegraphics[width=1\columnwidth]{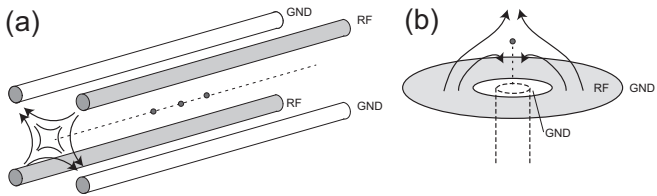}
\caption{Comparison of the traditional four-rod linear Paul trap (a) to the point Paul trap (b). The latter achieves quadrupole ion confinement through rf on a single, ring-shaped electrode. Dashed lines suggest how cylindrical elements -- such as optical fibers -- may be introduced to the point Paul geometry.\label{fig:fourRodCompare}}
\end{figure}

Traditionally, such devices have been rather bulky, three-dimensional structures that required precise machining and careful assembly. Recently, however, the four-rod linear Paul trap (see Fig.~\ref{fig:fourRodCompare}a) has been transformed into a two-dimensional structure, with all electrodes in a single plane above which ions can be trapped~\cite{Chiaverini2005a}. This new class of so-called surface traps offer a tremendous advantage over their predecessors in that electrodes can be defined lithographically with extremely high precision and that construction can leverage the techniques of microfabrication, with the possibility of incorporating the technology of CMOS for integrated control hardware~\cite{Kim2005,Leibrandt2009}. These aspects are particularly attractive to applications in  quantum information processing where limitations are currently, by a large degree, pertaining to the scalability of devices for trapping as well as certain elements of infrastructure such as optics, laser light delivery, and control electronics. 

In this paper we study a novel type of rf surface trap with a high degree of symmetry in its electrode geometry. The generic geometry of this trap, which we shall refer to as the point Paul trap, is shown in Fig.~\ref{fig:fourRodCompare}b, and may consist of any number of concentric electrodes of arbitrary widths to which different voltages can be applied. This design originated in a study of surface electrode traps \cite{Pearson2006,PearsonThesis2006}, but was subsequently strongly inspired by work on planar Penning traps~\cite{Stahl2005}, where a similar geometry was used to create a static electric quadrupole field that, when combined with a strong homogeneous magnetic field, gave rise to a confining potential above the surface of the electrodes. The point Paul trap also bears close resemblance to the rf ring- and the rf hole-traps~\cite{Brewer1992}; however differs in that the ion is trapped above the surface of the electrodes as opposed to in-between, which makes this geometry better suited for microfabrication. 

A consequence of the azimuthal symmetry of the electrodes is that the rf-field exhibits a nodal point rather that a nodal line as in the linear Paul trap and that the confining fields originate exclusively from the rf-potential, rendering the addition of dc-potentials nonessential for anything but the compensation of stray charges on the trap. 	This makes the point Paul trap well-suited for confinement of single ions, which may then reside at the rf-nodal point where the amplitude of the rapidly oscillating rf-field vanishes. 

The ability to fabricate these traps in a scalable fashion makes them attractive for realizing large arrays of single ions in independent traps that may be utilized for a quantum processor, provided the individual ions can be interconnected, e.g., through optical fibers~\cite{Olmschenk2009,Duan2010}. On this aspect, the axial symmetry of the trap lends itself well to integration of such fibers and potentially other optical elements that also possess axial symmetry. The fiber, for instance, may be introduced through the electrodes directly beneath the ions with minimal perturbation of the trapping fields.

Another possible application of this trap is in the field of quantum simulation. While classical computers are unable to efficiently simulate coupled spin systems, such simulations may be implemented using a quantum mechanical system of effective spins, such as a two-dimensional lattice of interacting ions. The resulting potential of the point Paul trap provides ion crystals with exactly the requisite two-dimensional planar structure. As such, the system could be used to simulate e.g. a frustrated spin system~\cite{Porras2004,Clark2009}, as was demonstrated recently \cite{Kim2010}.

We also find that our trap design is ideally suited for realizing a scheme by which the height of a single trapped ion above the trap surface is varied \emph{in-situ}. This capability may prove extremely useful in the search for the origin of anomalous heating in ion traps -- a problem currently impeding the advancement of quantum computation with trapped ions~\cite{Deslauriers2006,Labaziewicz2008}. It also provides a general technique by which oven contamination of the trap can be minimized by loading further away from the trap surface, and subsequently bringing the ion to the desired trap height.

This paper is organized as follows: in Section~\ref{Theory} we present a model for the planar point Paul trap, derive analytic expressions for the relevant trapping parameters, present full numerical results of trap optimization, and consider a scheme for the variation of the ion height above the trap surface. In Section~\ref{Setup} we describe our experimental setup used for the verification of the model; and in Section \ref{Results} we present experimental results for trapping of single and few ions in a printed circuit board (PCB) implementation of the point Paul trap.

\section{Point Paul trap model}\label{Theory}
We proceed with a general treatment applicable to an arbitrary number of circular electrodes, and then focus on a particular geometry that we will study experimentally later in Sections~\ref{Setup} and \ref{Results}. At the end of this section we also describe a scheme for variation of the ion height above the trap surface.

\subsection{Potential from annular planar electrodes}
\begin{figure}
\includegraphics[width=1\columnwidth]{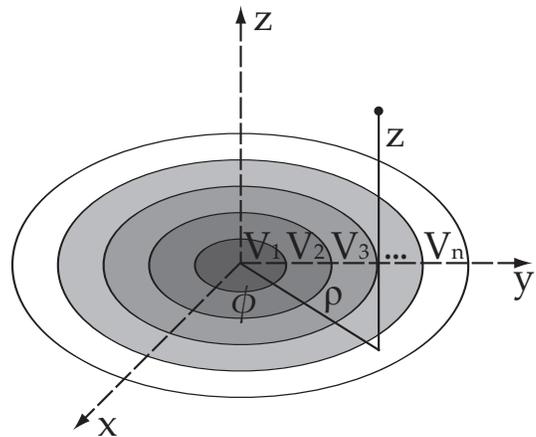}
\caption{The generic layout of the point Paul trap, which consists of concentric annular electrodes with arbitrary widths, illustrated in a cylindrical coordinate system.\label{fig:ElectrodesGeneric}}
\end{figure}
We begin with the general solution to the Laplace equation in charge-free space and express this in cylindrical coordinates for $z\geq0$, yielding~\cite{Jackson1999}
\begin{eqnarray}\label{eq:PotGeneral}
\Phi(z,\rho,\phi)&=&\sum_{m=0}^\infty \int_0^\infty J_m(k\rho) e^{-kz}\\  
&\times& \left[\mathcal{A}_m(k)\cos{(m \phi)}+\mathcal{B}_m(k)\sin{(m\phi)} \right] dk. \nonumber
\end{eqnarray}
where $J_m(k\rho)$ are the usual Bessel functions and $\mathcal{A}_m(k)$ and $\mathcal{B}_m(k)$ are coefficients to be determined based on the boundary conditions of the problem. Based on the azimuthal symmetry of Fig.~\ref{fig:ElectrodesGeneric}, Eq.~(\ref{eq:PotGeneral}) further simplifies to~\cite{Jackson1999}
\begin{equation}\label{eq:PotGeneral2}
\Phi(z,\rho)=\int_0^\infty J_0(k\rho) e^{-kz} \mathcal{A}_0(k) dk.
\end{equation}
In turn, $\mathcal{A}_0$ can be expressed as $\mathcal{A}_0(k)=\sum_i^nA_i(k)$, where
\begin{equation}\label{eq:Ai1}
A_i(k)=k\int_{\alpha_i}^{\beta_i}  \rho J_0(k\rho) V_i(\rho) d\rho. 
\end{equation}
All information about the electrode geometry is now included in the $\mathcal{A}_0$ coefficient, which we have in turn written as a sum of $n$ sub-coefficients, each accounting for the effect of a single annular electrode $i$ with inner radius $\alpha_i$, outer radius $\beta_i$ and a voltage $V_i$, which we shall assume is constant across the electrode. The integral of Eq.~(\ref{eq:Ai1}) can be evaluated using the identity for the Bessel functions $\int_0^u v J_0(v) dv=uJ_1(u)$ to give
\begin{equation}\label{eq:Ai2}
A_i(k)=V_i\left[\beta_i J_1(k\beta_i)-\alpha_i J_1(k\alpha_i)\right].
\end{equation}
This completes the general treatment of the problem: The electric potential above a surface at $z=0$ with $n$ concentric circular electrodes, each with independent voltages $V_i$ and inner and outer radii of $\alpha_i$ and $\beta_i$ respectively, is given by Eq.~(\ref{eq:PotGeneral2}) with $\mathcal{A}_0(k)=\sum_i^nA_i(k)$, where the $A_i$ coefficients are given by Eq.~(\ref{eq:Ai2}). 

\subsection{The three-electrode point Paul trap}\label{sec:AnalyticModel}
While static potentials alone may provide confinement in one or two dimensions, Earnshaw's theorem dictates that this is accompanied by a defocusing effect in the orthogonal dimensions. In the work of Ref.~\cite{Stahl2005}, three-dimensional confinement was achieved via the addition of a static magnetic field to realize a planar Penning trap. Here, we use a time-varying rf field to achieve charge confinement as a planar Paul trap. Namely, we consider the simple geometry of only three electrodes defined by the following boundary conditions:
\begin{equation}
\Phi(z=0,\rho)= 
\begin{cases} 
0  & \text{for $0<\rho<a$,}
\\
V_\mathrm{rf}\cos{(\Omega_\mathrm{rf}t)} & \text{for $a\leq \rho \leq b$,}
\\
0 & \text{for $b<\rho<\infty$,}
\end{cases}
\end{equation}
where $V_\mathrm{rf}$ is the amplitude of the applied voltage and $\Omega_\mathrm{rf}$ is the frequency. The resulting potential then reads
\begin{equation}\label{eq:PotGeneral3}
\Phi(z,\rho,t)=V_\mathrm{rf}\cos{(\Omega_\mathrm{rf}t)} \kappa(z,\rho),
\end{equation} 
where
\begin{equation}\label{eq:kappa}
\kappa(z,\rho)=\int_0^\infty  \left[ b J_1(kb)-a J_1(ka) \right] e^{-kz} J_0(k\rho) dk.
\end{equation} 

In general, Eq.~(\ref{eq:kappa}) has to be solved numerically. However, for the case of $\rho=0$ the problem simplifies significantly and an analytic solution can be obtained. From the symmetry of the problem we can infer that if a nontrivial field zero ($\vec{E}=0$) exists for $z>0$, it will be located on the axis defined by $\rho=0$ and hence this scenario is worthy of attention. 

The on-axis potential is easily integrated to yield:
\begin{equation}\label{eq:kappa_z}
\kappa(z,0) = \frac{1}{\sqrt{1+(\frac{a}{z})^2}}-\frac{1}{\sqrt{1+(\frac{b}{z})^2}}.
\end{equation}
Asserting that this potential provides for an electric field null at $z=z_0>0$, we expand $\Phi(z,t)$ around this point and use the resulting expansion to find the equation of motion for a particle of mass $M$ and charge $Q$ along the $z$-axis:
\begin{eqnarray}\label{eq:d2z}
\ddot{z}(t)&=&-\frac{Q}{M}\frac{\partial \Phi(z,t)}{\partial z} \\
&\simeq&-\frac{QV_\mathrm{rf}}{M}\cos{(\Omega_\mathrm{rf}t)} \times \left[f(a,b) (z-z_0) + O(z-z_0)^2 \right].\nonumber
\end{eqnarray}
Provided $|z-z_0|\ll z_0$, which is a reasonable assumption for a trapped ion, terms of second and higher order can be neglected, and the equation of motion takes the form of the well-known Mathieu equation. With proper rescaling of variables, Eq.~(\ref{eq:d2z}) can be cast into the standard Mathieu form~\cite{Abramowitz1964}:
\begin{equation}\label{eq:Mathieu}
\ddot{\tilde{z}}(\tau)+2q\cos{(2\tau)}\tilde{z}(\tau)=0,
\end{equation}
where we have substituted $\tilde{z}=z-z_0$ and $\tau=\Omega_\mathrm{rf}t/2$, and the Mathieu q-parameter has been defined as
\begin{equation}\label{eq:Mathieu_q}
q=\frac{2QV_\mathrm{rf}}{M\Omega_\mathrm{rf}^2}f(a,b).
\end{equation}
Here, everything related to the trap geometry is collected into a single function $f(a,b)$ of unit $\left[\mathrm{length}\right]^{-2}$ given by
\begin{equation}\label{eq:f}
f(a,b)=\sqrt{\frac{9 (b^\frac{2}{3} - a^\frac{2}{3})^2 (b^\frac{2}{3} + a^\frac{2}{3})^6}{b^\frac{4}{3} a^\frac{4}{3} (b^\frac{4}{3} +b^\frac{2}{3}a^\frac{2}{3} + a^\frac{4}{3})^5}}.
\end{equation}
Note that in this treatment $a$ denotes the inner radius of the rf electrode and not the Mathieu $a$-parameter, commonly used in the literature on Paul traps. The Mathieu $a$-parameter, which corresponds to the inclusion of a dc-potential in the equation of motion (Eq. \ref{eq:d2z}), is rendered superfluous in the point Paul trap as full three-dimensional confinement is achieved in this geometry by the rf-field alone.

When the trap is operated such that $|q|\ll1$, the equation of motion can be readily solved to yield
\begin{equation}\label{eq:MathieuSol}
\tilde{z}(t)=\sigma_0\left[1-\frac{q}{2}\cos{(\Omega_\mathrm{rf}t)}\right]\cos{(\omega_z t)}.
\end{equation}
This is the usual result, familiar also from the four-rod linear Paul trap, where the motion is comprised of two distinct types of motion: a slow, so-called \emph{secular}, motion with an amplitude $\sigma_0$ at a frequency
\begin{equation}\label{eq:secular_freq_z}
\omega_z = \frac{q}{2\sqrt{2}}\Omega_\mathrm{rf}=\frac{QV_\mathrm{rf}}{\sqrt{2}M\Omega_\mathrm{rf}}f(a,b)
\end{equation}
and a superimposed, fast \emph{micromotion}, with a lower amplitude of $\frac{1}{2}q\sigma_0$ and at sideband frequencies of the rf drive $\Omega_\mathrm{rf}$. Neglecting the micromotion -- a reasonable approximation for $|q|\ll1$ -- we can define an approximate harmonic potential that describes the ion motion near the quadrupole zero by the following
\begin{equation}\label{eq:PseudoPot2}
\Psi(z)=\frac{1}{2}M\omega_z^2(z-z_0)^2=\frac{Q^2V_\mathrm{rf}^2}{4 M\Omega_\mathrm{rf}^2}f^2(a,b)(z-z_0)^2,
\end{equation}
thereby showing the charge-confining capabilities of the three-electrode point Paul trap, provided that $f^2>0$.

\subsection{Trap optimization and results}\label{sec:TrapOpt}
While the harmonic potential of Eq.~(\ref{eq:PseudoPot2}) provides an intuitive connection to the physical, time-averaged motion of the trapped ion in the vicinity of the rf node, it does not reveal any information about the dynamics where the inequality $|z-z_0|\ll z_0$ is not satisfied. For instance, in a real device there is necessarily a finite trapping volume and trap depth. These quantities originate from the shape of the potential significantly beyond the harmonic region. In the limit $q\ll 1$, the effective potential energy beyond the harmonic regime -- commonly referred to as the \emph{pseudopotential} -- may be expressed directly through the gradient of the electric potential, here written in terms of $\kappa$, as~\cite{Dehmelt1967,Ghosh1995,Wesenberg2008}
\begin{equation}\label{eq:PseudoPotAlt}
\Psi(z,\rho) = \frac{Q^2V_\mathrm{rf}^2}{4M\Omega_\mathrm{rf}^2}\left|\nabla \kappa(z,\rho)\right|^2.
\end{equation}
\begin{figure}
\includegraphics[width=1\columnwidth]{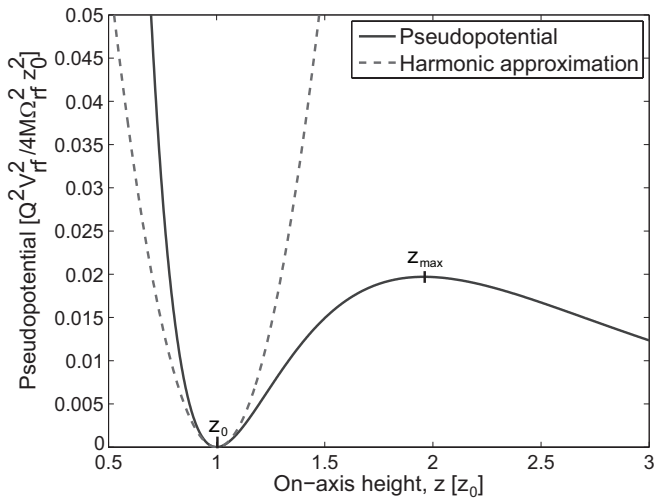}
\caption{The harmonic potential approximation (Eq. \ref{eq:PseudoPot2}; dashed line) for the point Paul trap is shown on top of the pseudopotential (Eq. \ref{eq:PseudoPotAlt}; solid line). The latter identifies the trap turn-around point $z_\mathrm{max}$ in addition to the trap location $z_0$. Trap geometry $(a,b)$ is chosen as in Table \ref{table:optTrap}.\label{fig:Pseudopotential}}
\end{figure}
The solid line in Fig.~\ref{fig:Pseudopotential} shows the pseudopotential for the planar point Paul trap. Superimposed is the harmonic approximation (dashed line). Inserting the expression for $\kappa$ from Eq.~(\ref{eq:kappa}) into the expression for the pseudopotential (Eq. \ref{eq:PseudoPotAlt}) yields two physically meaningful extrema,
\begin{equation}\label{eq:z0}
z_0=\sqrt{\frac{b^{4/3} a^{4/3}}{b^{2/3} + a^{2/3}}}
\end{equation}
and
\begin{equation}\label{eq:z1}
z_\mathrm{max} = \sqrt{\frac{b^{6/5} - a^{6/5}}{a^{-4/5} - b^{-4/5}}}.
\end{equation}
Here, $z_0$ is the coordinate of the pseudopotential minimum, and $z_\mathrm{max}$ denotes the turning-point of the confining pseudopotential. The difference $z_\mathrm{max}-z_0$ can be taken as a linear measure of the effective trap volume. Furthermore, the corresponding trap depth can now be defined as $D=\Psi(z_\mathrm{max})-\Psi(z_0)$. Using Eq.~(\ref{eq:z0}) and Eq.~(\ref{eq:z1}), one finds that the trap depth is positive for all values of $b>a>0$, and is equivalent to: 
\begin{equation}\label{eq:D}
D= \frac{Q^2V_\mathrm{rf}^2}{4M\Omega_\mathrm{rf}^2}\left[a^2(a^2 + z_\mathrm{max}^2)^{-\frac{3}{2}} - b^2(b^2 + z_\mathrm{max}^2)^{-\frac{3}{2}}\right]^2.
\end{equation}

The trap depth is a reasonable quantity to be optimized in the design of the point Paul trap. However, unconstrained optimization of Eq. (\ref{eq:D}) over $(a,b)$ will influence not only the trap depth but also the trap height $z_0$ above the surface through Eq.~(\ref{eq:z0}). Often in experiments, the trap height is a parameter of importance and so a more reasonable strategy is to optimize the trap depth for a fixed value of $z_0$. This can in principle be done analytically; however, the results are more useful in their numerical form. Table~\ref{table:optTrap} summarizes the results of this optimization. For the purpose of comparison with the four-rod linear Paul trap we have defined 
\begin{equation}\label{eq:linPaulTrap}
q_\mathrm{4rod}\equiv \frac{2QV_\mathrm{rf}}{M\Omega_\mathrm{rf}^2z_0^2}~\quad \mathrm{and}~\quad~D_\mathrm{4rod}\equiv \frac{Q^2V_\mathrm{rf}^2}{4M\Omega_\mathrm{rf}^2z_0^2},
\end{equation}
which corresponds to the $q$-parameter and the trap depth, respectively, for the three-dimensional four-rod Paul trap with an ion-electrode distance of $z_0$.
\begin{table}
 \caption{Results of numerical optimization of the trap depth for a fixed trap height $z_0$. The choice of units allows for direct comparison with the three-dimensional linear Paul trap.\label{table:optTrap}}
 \begin{ruledtabular}
 \begin{tabular}{ccccc}
 $a \left[z_0\right]$ & $b \left[z_0\right]$ & $z_\mathrm{max} \left[z_0\right]$ & $q \left[q_\mathrm{4rod}\right]$ & $D \left[D_\mathrm{4rod}\right]$\\
$0.651679$ & $3.57668$ & $1.957965$ & $ 0.471565$ & $0.019703$\\
 \end{tabular}
 \end{ruledtabular}
 \end{table}

From the optimization results of Table~\ref{table:optTrap}, it is seen that the $q$-parameter of the point Paul trap is roughly a factor $\frac{1}{2}$ of the four-rod linear Paul trap while the trap depth is about a factor $\frac{1}{50}$. By comparison, the surface electrode linear Paul trap design that has recently attracted much attention in the context of quantum computing~\cite{Seidelin2006,Wang2010} has a $q$-parameter and a trap depth that are approximately a factor $\frac{1}{2\sqrt{3}}$ and $\frac{1}{72}$, respectively, of the four-rod linear Paul trap~\cite{Chiaverini2005a}.

\subsection{3D potential of the point Paul trap}
\begin{figure}
\includegraphics[width=1\columnwidth]{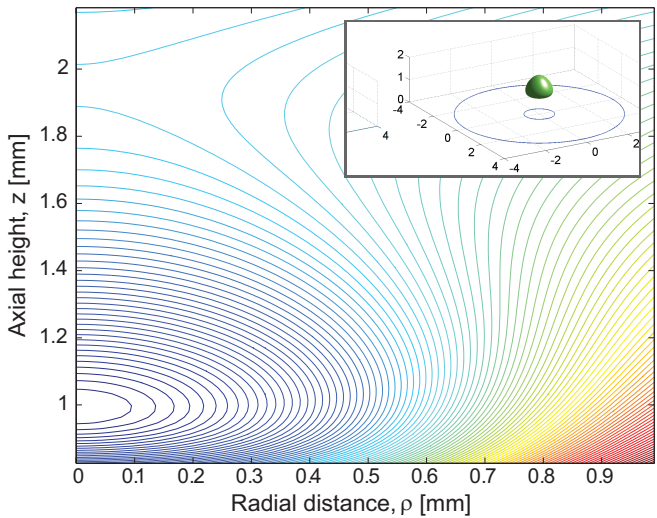}
\caption{(Color online) A contour plot for the pseudopotential (Eq.~(\ref{eq:PseudoPotAlt})) of a $z_0=1$mm trap, using the optimal geometry of Tabe \ref{table:optTrap}. Inset shows the trap isosurface in relation to outlines of the rf ring electrode (units in $z_0$).\label{fig:3Dpot}}
\end{figure}

To extract information about the three-dimensional shape of the pseudopotential, we insert the full expression of Eq.~(\ref{eq:kappa}) into Eq.~(\ref{eq:PseudoPotAlt}) and integrate numerically at discrete values of $\rho$ and $z$. This yields a contour plot as shown in Fig.~\ref{fig:3Dpot}, where the trap dimensions are chosen according to Table \ref{table:optTrap} for a trap height of $z_0=1$~mm. It is seen that the confinement is tighter along the axial direction of the trap than along the radial direction. Also shown (see inset) is the isosurface corresponding to the edge of the trap. 

A unique feature of this trap design is that confinement is achieved in three dimensions using only an rf-field as opposed to the linear trap designs that also require static dc-potentials along the axis defined by the nodal line of the rf quadrupole field. One consequence of this is that, similar to the four-rod linear Paul trap but in contrast to the linear surface electrode Paul trap, the point Paul trap is naturally compensated and dc-potentials are in principle not required for stable trapping.

Another implication of achieving three-dimensional confinement with a single rf-field is that the ratio of the axial and radial secular frequencies is fixed by the geometry of the trap. From quadratic fits to the central harmonic region of Fig.~\ref{fig:3Dpot}, the ratio of the radial to the axial trap frequency is found to be $\frac{\omega_\rho}{\omega_z}\simeq 0.50$.

\subsection{Scheme for variation of the nodal point of the rf-field}
We mentioned previously that the ion height above the trap surface is often a parameter of interest and designs are generally optimized with some focus on this parameter. Controlling the ion position \emph{in-situ} is a desirable capability for a variety of experimental applications. To mention a few examples, in experiments incorporating optical cavities, careful alignment of the cavity with respect to the ion is necessary to achieve the highest possible coupling between the ion and the cavity mode~\cite{Keller2004,Herskind2009a,Leibrandt2009a,Dubin2010}; and in the study on anomalous heating in ion traps the ion height is a key parameter, with the scaling of the heating rate currently believed to follow a $z_0^{-4}$ trend~\cite{Deslauriers2006}. The ion position may be adjusted via the addition of dc-potentials; however, unless the rf quadrupole potential is adjusted accordingly, the ion location will not coincide with the zero of the rf field and as a result its motion will be driven by the rapidly changing rf-fields, resulting in broadening of the atomic transitions~\cite{Berkeland1998}. It is possible to align the trap relative to external objects such as mirrors by physically moving the trap~\cite{Keller2004,Leibrandt2009a,Dubin2010}; however, recently a method for translating the node of the rf quadrupole field has been developed and used both in four-rod Paul traps~\cite{Herskind2009} and in planar surface traps~\cite{Cetina2007,VanDevender2010} to shift the ions without introducing excess micromotion.

The basic principle of this method is to apply different amplitudes of rf-potential on individual electrodes, thus causing a shift of the rf field node relative to those electrodes. Implementation of this technique is particularly simple in our geometry where it is achieved by adding an rf-field to the central, previously grounded, electrode. The resulting boundary conditions for the electric potential then becomes
\begin{equation}\label{eq:BoundaryCond2}
\Phi(\rho,0)= 
\begin{cases} 
\epsilon V_\mathrm{rf}\cos{(\Omega_\mathrm{rf}t+\theta)}  & \text{for $0<\rho<a$,}
\\
V_\mathrm{rf}\cos{(\Omega_\mathrm{rf}t)} & \text{for $a\leq \rho \leq b$,}
\\
0 & \text{for $b<\rho<\infty$,}
\end{cases}
\end{equation}
where $\epsilon$ and $\theta$ accounts for the amplitude and phase difference between the inner and the outer rf-electrodes. 

Before proceeding to find a solution to the potential in this configuration, it is useful analyze the scenario in qualitative terms to gain some intuition about the influence of this second rf-potential:

Along the $z$-axis, the rf-field from the outer ring electrode reverses its sign around the quadrupole zero point $z_0$, while the field from the inner rf-electrode is always in the same direction on the $z$-axis, pointing away from the trap surface. If the two electrodes are driven in-phase, their fields will be of opposite sign for $z<z_0$ and same sign for $z>z_0$. In that case the effect of the second rf-field is to decrease the magnitude of the field below the original trap location and increase it above, thus bringing the rf node closer to the electrodes. Similarly, if the two rf-electrodes are driven out-of-phase the trapping point will move away from the electrode surface.

To develop a quantitative model, we again solve the potential for the boundary conditions of Eq. (\ref{eq:BoundaryCond2}), but for the case of either in-phase ($\theta=0$) or out-of-phase ($\theta=\pi$) drives. The scenario where the two electrodes are related by some other relative phase should be avoided, as it will result in excess micromotion, analogously to the case of the four-rod linear Paul trap~\cite{Berkeland1998}. Absorbing the phase into the sign of $\epsilon$, the spatial part of the resulting potential reads 
\begin{eqnarray}\label{eq:kappa_var_rf}
\kappa(z,\rho)&=&\int_0^\infty dk e^{-kz} J_0(k\rho)\\
&\times& \left[ b J_1(kb)-(1-\epsilon)a J_1(ka)\right]. \nonumber
\end{eqnarray}  
We again focus on the case of $\rho=0$ to find an analytic expression for $\kappa(z,0)$:
\begin{equation}\label{eq:kappa_z_var_rf}
\kappa(z,0)= 
\frac{1}{\sqrt{1+(\frac{a}{z})^2}}-\frac{1}{\sqrt{1+(\frac{b}{z})^2}}+\epsilon\left(1-\frac{1}{\sqrt{1+(\frac{a}{z})^2}}\right),
\end{equation}
where $\epsilon>0$ and $\epsilon<0$ correspond to the in-phase and out-of-phase drives, respectively.

\begin{figure}
\includegraphics[width=1\columnwidth]{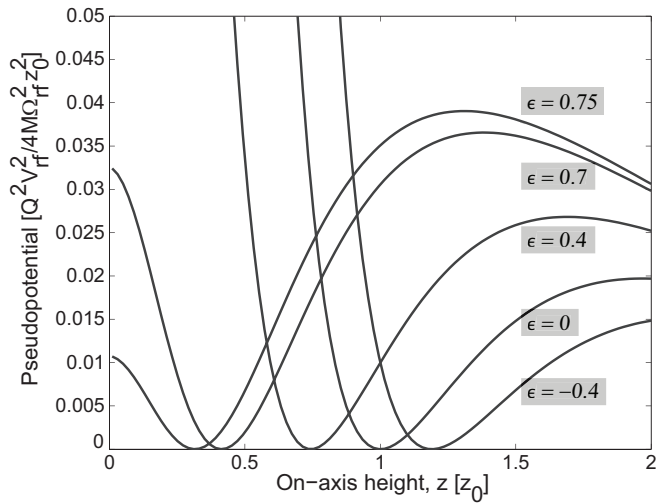}
\caption{Variation in the on-axis pseudopotential $\Psi$ with the addition of secondary rf at various values of $\epsilon$. The trap dimensions $a$ and $b$ are as found in Table \ref{table:optTrap}.\label{fig:HeightVarInOutBoth}}
\end{figure}
Inserting the expression for $\kappa(z)$ into Eq.~(\ref{eq:PseudoPotAlt}) and using the geometry of Table~(\ref{table:optTrap}) we may plot the pseudopotential for various values of $\epsilon$. Figure~\ref{fig:HeightVarInOutBoth} shows such example plots for various values of $\epsilon$. It is seen that, in accordance with the qualitative model, the in-phase drive ($\epsilon>0$) lowers the trap height $z_0$, while the out-of-phase drive ($\epsilon<0$) increases the height.

The new trap location and turning point can be found analytically as in Section~\ref{sec:TrapOpt}, yielding:
\begin{equation}\label{eq:z0HeightVar}
z'_0(\epsilon)=\sqrt{\frac{b^2a^{4/3}(1-\epsilon)^{2/3}-a^2b^{4/3}}{b^{4/3}-a^{4/3}(1-\epsilon)^{2/3}}}
\end{equation}
and
\begin{equation}\label{eq:z0HeightVarMax}
z'_\mathrm{max}(\epsilon)=\sqrt{\frac{b^2a^{4/5}(1-\epsilon)^{2/5}-a^2b^{4/5}}{b^{4/5}-a^{4/5}(1-\epsilon)^{2/5}}}
\end{equation}
We also find an expression for the $q$-parameter, which is derived analogously to Eq.~(\ref{eq:Mathieu_q}) using (\ref{eq:kappa_z_var_rf}):
\begin{equation}\label{eq:Mathieu_q_HeightVar}
q'(\epsilon)=\frac{2QV_\mathrm{rf}}{M\Omega_\mathrm{rf}^2}\left[\frac{3a^2z'_0(1-\epsilon)}{(a^2+z_0'^2)^{5/2}}-\frac{3b^2z'_0}{(b^2+z_0'^2)^{5/2}}\right].
\end{equation}

\begin{figure}
\includegraphics[width=1\columnwidth]{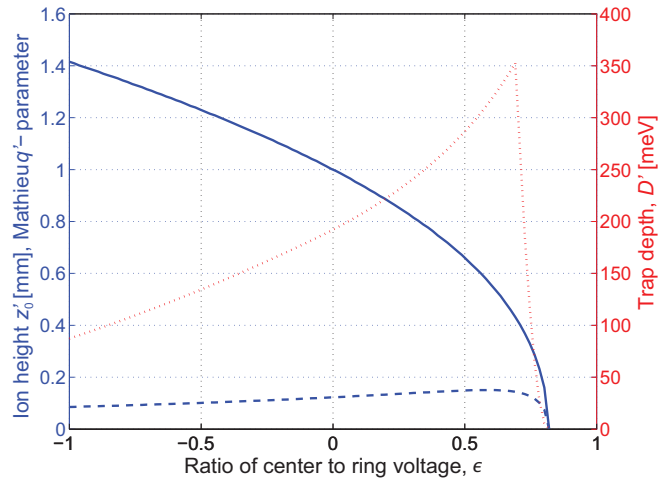}
\caption{(Color online) The variation in trap height (solid blue line), Mathieu $q$-parameter (dashed blue line) and the effective trap depth (dotted red line) as a function of the amplitude of the secondary rf. These parameters are computed for an optimal $z_0=1$mm (under single-rf) $^{88}$Sr$^+$ trap, with $V_\mathrm{rf}=300$V and $\Omega_\mathrm{rf}=2\pi\cdot8$MHz.\label{fig:HeightVar_z0}}
\end{figure}

In addition to the ion height variation, the presence of a second rf modifies the overall shape of the pseudopotential and hence also the effective trap depth, as evidenced already in Fig.~\ref{fig:HeightVarInOutBoth}. Namely, the out-of-phase regime is limited by a diminishing barrier on the side further away from the trap, and the in-phase drive causes a lowering of depth on the side towards the plane. For an optimal $z_0=1$mm trap for $^{88}$Sr$^+$, Fig.~\ref{fig:HeightVar_z0} summarizes the variation in ion height $z'_0$, the Mathieu $q'$-parameter, and overall trap depth $D'$ as a function of $\epsilon$ under typical operating parameters. In particular, note the cusp in trap depth at $\epsilon \approx 0.7$, due to the rapidly diminishing trap barrier on the side towards the trap. Conservatively, a range of $0<\epsilon<0.7$ leads to a dynamic range of about $0.6 z_0 = 600\mu$m that is achievable without suffering any decrease in trap depth from the single-rf configuration. Of course, a reduction in the trap depth due to this technique may be compensated by varying the magnitude of the rf voltage and the rf frequency.

Compared with previous work on shifting the rf nodal line of a four-rod linear Paul trap~\cite{Herskind2009}, this range is a significant increase. The underlying reason is that the geometry of the point Paul trap is more favorable for such a scheme in that a shift in the ion height does not change the symmetry of the trap axis with respect to the electrodes, in contrast to the implementation in the four-rod linear Paul trap~\cite{Herskind2009} but similar to recent work on a related surface electrode ion trap~\cite{VanDevender2010}. 

The ability of the point Paul trap to vary the ion height without incurring micromotion would be of tremendous value in the search for the origin of anomalous heating in ion traps~\cite{Deslauriers2006,Labaziewicz2008}. Previous work on this problem have either used a technically challenging setup in which the trap electrodes were moved \emph{in-situ}~\cite{Deslauriers2006}, or has relied on systematic testing of traps of identical geometry but different scales, making this method prone to random errors associated with the fabrication of the individual traps~\cite{Labaziewicz2008}. 

Anomalous heating is believed to originate from fluctuating patch potentials on the electrodes. The model describing this effect predicts a scaling of the heating rate of $1/z_0^4$; however, only one experiment has thus far been able to conduct a systematic study to confirm this in a single trap geometry~\cite{Deslauriers2006}. As the suggested scheme for the point Paul trap in principle allows for  \emph{in-situ} variation of the ion height by almost an order of magnitude (with modest adjustments in drive rf), it provides for an extremely sensitive test of the patch potential model without complications associated with physically moving the trap electrodes, and without the errors and difficulty in obtaining good statistics associated with the use of individual traps for each value of $z_0$.

\section{Experimental setup}\label{Setup}

\begin{figure}
\includegraphics[width=1\columnwidth]{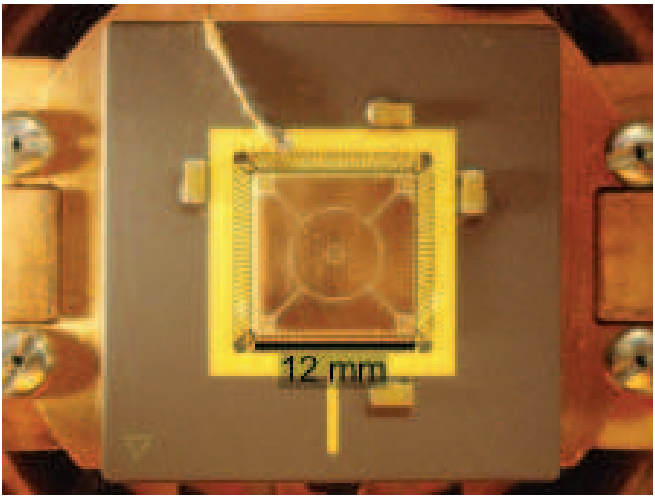}
\caption{(Color online) Image of the PCB trap mounted on the CPGA and installed on the 8K baseplate of the closed-cycle cryostat. Capacitors are connected to the DC electrodes in order to minimize rf pickup.\label{fig:pcbTrap}}
\end{figure}
To validate the model presented in the previous section we have tested a planar point Paul trap with an electrode geometry similar to that of section~\ref{sec:AnalyticModel}, and characterized its trapping of $^{88}$Sr$^+$ ions. The trap itself is based on a printed circuit board (PCB) with copper electrodes on a low-rf-loss substrate (Rogers 4350B, fabricated by Hughes circuits)~\cite{Brown2007}. Figure~\ref{fig:pcbTrap}  shows a picture of this trap mounted in a ceramic pin grid array (CPGA) chip carrier. The radius of the inner ground electrode is $a_1=550\mu$m, the inner radius of the rf ring electrode is $a_2=650\mu$m and its outer radius is $b=3.24$mm. Due to the $100\mu$m gap between the electrodes, the ratio $\frac{a_{1,2}}{b}$ deviates by about $10\%$ from the optimum geometry of Table~\ref{table:optTrap}, and produces in an ion height of $z_0=960\mu$m using $a_2$. 

In this particular realization of the model system of Fig.~\ref{fig:ElectrodesGeneric}, the outer ground has been segmented into four electrodes and their independent potentials can each be adjusted to compensate for stray electric fields in the vicinity of the trap. Boundary element analysis of the exact electrode configuration predicts an ion height of $940\mu$m, in agreement with our analytical model. This analysis also finds the ratio of radial to axial secular frequencies to be $\frac{\omega_\rho}{\omega_z}\simeq0.50$, again in good agreement with the analytic result.

The trap is mounted on the 8K baseplate of a close-cycle cryostat described in Ref.~\cite{Antohi2009}. Ions are loaded via resonant photoionization of an atomic beam from an effusive oven that is heated resistively to a few hundred degrees Celcius during loading. Once ionized, the ions are Doppler-cooled using light at 422nm and 1092nm. The relevant level scheme is shown in the inset of Fig.~\ref{fig:data_telegraph} along with the lifetimes of the excited states. Typically, $20\mu$W of $422$nm light focused to a $50\mu$m waist at the location of the ion is used for the $5^2$S$_{1/2}\leftrightarrow 5^2$P$_{1/2}$ transition, while about $20\mu$W of $1092$nm focused to a $150\mu$m waist repumps the ion on the $4^2$D$_{3/2}\leftrightarrow 5^2$P$_{1/2}$ transition. Dark states of the D$_{3/2}$-manifold are destabilized by polarization modulation of the 1092nm light at 10~MHz as described in~\cite{Berkeland2002}. Fluorescence from the ions is collected with a NA$\simeq0.5$ lens mounted inside the vacuum chamber  and imaged onto a CCD camera (Andor Luca R) and a photomultiplier tube (PMT) (Hamamatsu, H7360-02), the latter achieving an overall detection efficiency (including losses on optics in the imaging system) of $0.6\%$. We ensure that the ions are located at the nodal point of the rf-field by minimizing the micromotion amplitude using the correlation measurement technique as described in Ref.~\cite{Berkeland1998}.

Radial and axial secular frequencies can be measured by applying a small sinusoidal voltage to a DC electrode, thus exciting motion along the corresponding direction~\cite{Naegerl1998}. At the resonant frequency of the trap, the strong increase in the amplitude of this driven motion is detected as a sudden change in the fluorescence from the ion, and can also be corroborated by increased motion along an axis by the imaging system. Depending on the trap depth, amplitude of the perturbative signal and the sweep rate, this allows a measurement of the secular frequencies with a typical accuracy of about $\pm 5$kHz.

Finally, the shift in ion height with the addition of a second rf drive is determined by the translation of the doppler cooling beam necessary in order to maximize ion fluorescence. The spatial position of the beam is calibrated by a digital translation stage with submicron resolution. However, the accuracy of this method is ultimately limited by the finite waist of the laser beams.

\section{Results}\label{Results}

\begin{figure}
\includegraphics[width=1\columnwidth]{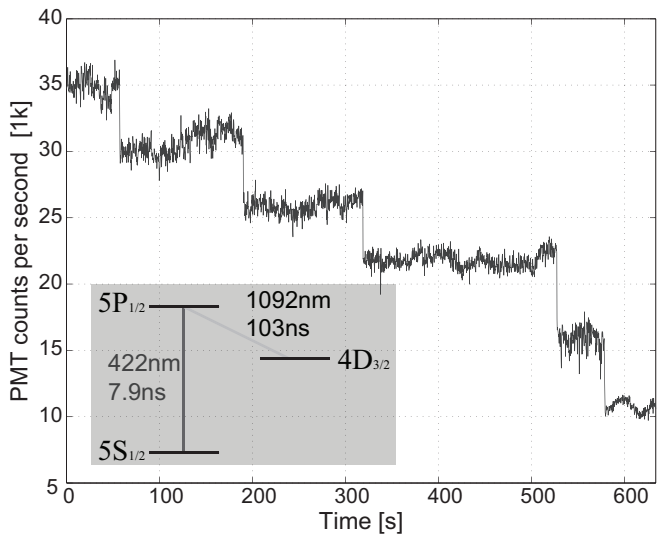}
\caption{Telegraph log shows the discrete loss of five ions from the PCB trap, as measured by scattered $422$nm light from the $5^2$S$_{1/2}\leftrightarrow 5^2$P$_{1/2}$ transition (see inset).\label{fig:data_telegraph}}
\end{figure}

The purpose of this section is to validate the model presented in Section~\ref{Theory}. Fig.~\ref{fig:data_telegraph} shows a time log of the fluorescence signal detected with the PMT as five ions are loaded and then lost from the trap. This rapid ion loss was observed prior to compensation of the stray dc fields and optimization of Doppler cooling. The discrete steps in photon counts provides a calibration of the scale to distinguish a single ion from two or more. Once compensation is optimized, we find that the trapping is stable for several hours when Doppler cooling is applied, limited only by the long-term stability of our lasers.

\begin{figure}
\includegraphics[width=1\columnwidth]{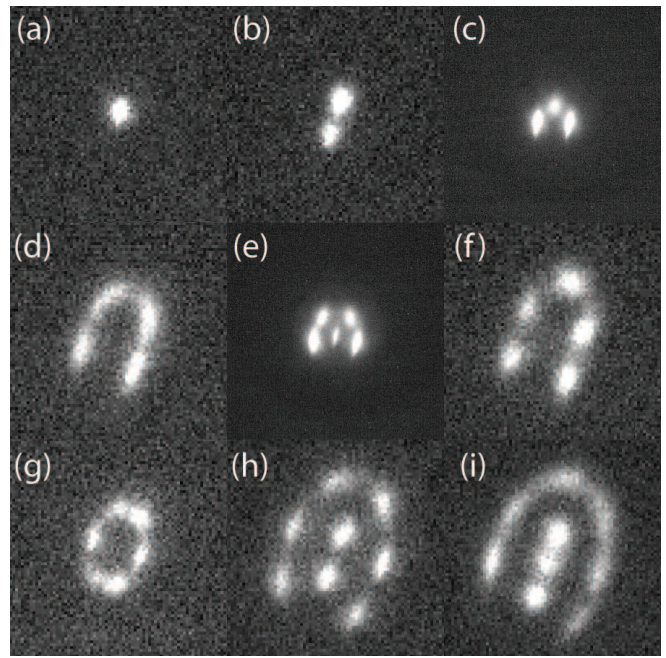}
\caption{Summary of the ion crystals that were observed with the PCB trap. The crystals in panels (c),(e) were observed at an ion height of $940\mu$m ($V_\mathrm{rf}=360$V, $10$s exposure) where the panel's field of view is $70\mu$m$\times70\mu$m. The remaining crystals were observed at a height of about $600\mu$m ($V_\mathrm{rf}=275$V, $\epsilon\simeq0.52$, $500$ms exposure) where each panel corresponds to approximately $40\mu$m$\times40\mu$m.\label{fig:data_crystals}}
\end{figure}

In addition to the discrete PMT logs, we have been able to resolve individual ions in 2D crystals involving up to nine ions, which is summarized in Figure~\ref{fig:data_crystals}. As noted in the introduction, such lattices could be used for quantum simulations of, for instance, frustrated spin systems \cite{Clark2009,Kim2010}.

\begin{figure}
\includegraphics[width=1\columnwidth]{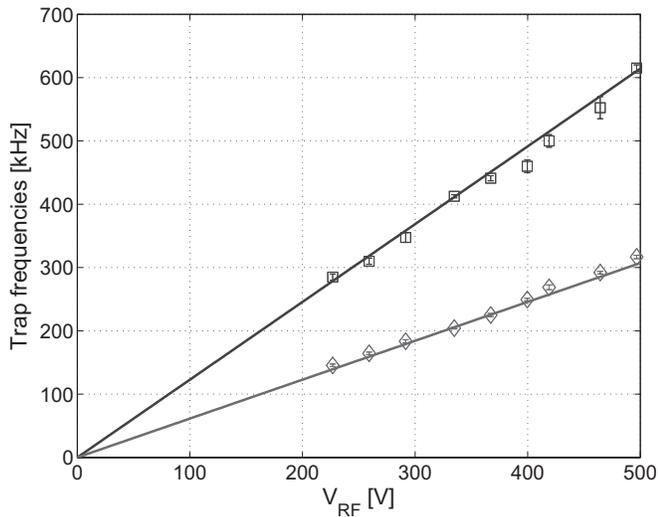}
\caption{Measurements of axial (square markers) and radial (diamond markers) secular frequencies of the point Paul trap. Data was taken at $z_0=940\mu$m under the single-rf drive at $\Omega_\mathrm{rf}=2\pi\times 8.07$MHz. Also shown are theoretically expected secular frequencies (not fits) according to Eq. (\ref{eq:secular_freq_z}).\label{fig:data_trapfreq}}
\end{figure}

The pseudopotential model for the point Paul trap is evaluated by measuring the secular frequencies of the trap for various applied rf-voltages $V_\mathrm{rf}$ at a constant rf-drive frequency of $\Omega_\mathrm{rf}=2\pi \times 8.07$MHz. The results are shown in Fig.~\ref{fig:data_trapfreq} for a single ion held at $~940\mu$m (using the numerical result) under the single-rf scheme. Solid lines are the theoretically expected trap frequencies (not fits) according to the model of Section \ref{Theory}, where the rf voltage has been calibrated independently by characterizing the driving helical resonator. Fits through the measured data yields a ratio of the radial to axial secular frequencies of $\frac{\omega_\rho}{\omega_z} = 0.51\pm 0.01$, in excellent agreement with the numerically predicted $\frac{\omega_\rho}{\omega_z} = 0.50$.

\begin{figure}
\includegraphics[width=1\columnwidth]{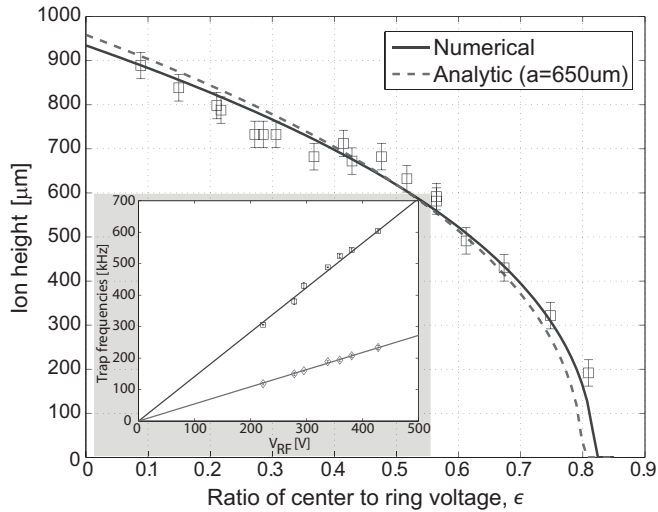}
\caption{Variation in ion height with the addition of the second rf on the innermost electrode. Solid curve shows the results of the numerical boundary element analysis, while the dashed curve shows the height as predicted by Eq. \ref{eq:z0HeightVarMax}. Inset: Measured axial (squares) and radial (diamonds) trap frequencies and their linear fits at an ion height of $600\mu$m.\label{fig:data_ionheight}}
\end{figure}

To establish the validity of the ion height variation model, we have explored the in-phase ($\epsilon>0$) parameter space. The implementation of a second in-phase drive was achieved by connecting a trim capacitor between the ring and center electrodes, which then formed a capacitive divider for the center electrode in combination with the intrinsic capacitance from the PCB. Figure~\ref{fig:data_ionheight} shows the results. Experimentally, we found that stable trapping was extremely straightforward to achieve, and single ions and crystals were trapped as close as $200\mu$m from the surface of the trap. Further approach was prevented by scattering of the incident laser beams from the trap surface, which interfered with ion detection. 

In addition, we were also able to measure secular frequencies when the ion was offset to a height of $600\mu$m, as shown in the inset of Fig.~\ref{fig:data_ionheight}. The comparable linearity in secular frequencies between the single- and dual-rf cases -- as well as the well-resolved ions of Fig.~\ref{fig:data_crystals} at an offset height -- suggests that the addition of a secondary rf has not added significant micromotion. The measured ratio of radial to axial secular frequencies was $\frac{\omega_\rho}{\omega_z}=0.38{\pm}0.01$, which deviates significantly from the expected $0.50$ (according to the model, the ratio remains unaffected by the addition of the second rf). Possible sources for deviation from the ideal model include the dc-potentials, which were added to ensure a well-compensated trap but which could have shifted secular frequencies. In particular, as the ion is brought closer to the trap surface, it is more susceptible to the effects of dc-potentials. Such fields may also break the degeneracy of the radial modes. We have in fact observed such separate radial modes, although in the case of the data in Fig.~\ref{fig:data_ionheight}, they were only separated by about $20$kHz, and thus average radial values are presented in that figure. We have numerically confirmed that the trap's dc potentials yield simultaneously a splitting of $20$kHz in the radial modes, and a modified secular frequency ratio of $0.40$.

\section{Conclusions}\label{Conclusions}

We have presented an analytic model of a circularly-symmetric rf surface trap and its experimental validation. This particular geometry offers several advantages for further investigations in quantum information processing. Firstly, the shape of the resulting potential leads to confinement of 2D ion crystals, which may be used as a platform for quantum simulation of interacting spins. Secondly, because the confinement is achieved through the ring electrode alone, the trap permits a straightforward scheme for variation of ion height \emph{in situ}. In this work, we have demonstrated almost an ion height variation over $200$--$1000\mu$m, which may then, for instance, be used for a stringent test of the supposed $z_0^{-4}$ scaling in anomalous ion heating. Finally, the axial symmetry of the trap lends itself to natural integration with optical fibers. Such a device, in turn, could serve as nodes in a quantum network, or provide an interface medium between ions and cold neutral atoms.
\begin{acknowledgments}
This work was supported by NSF CUA, and the COMMIT project funded by the ARO. THK was supported by the Siebel Scholar Foundation. PFH is grateful for the support from The Carlsberg Foundation and the Lundbeck Foundation.
\end{acknowledgments}

%

\end{document}